\documentclass[epj]{svjour}

\usepackage{exscale}
\usepackage[intlimits]{amsmath}
\usepackage{amsfonts}
\usepackage{amssymb,amscd}                
\usepackage{array}
\usepackage{wrapfig}
\usepackage{bbm}
\usepackage{multirow}
\usepackage[x11names]{xcolor}
\usepackage{graphicx}
\usepackage[T1]{fontenc}
\usepackage{subfigure}

\newcommand{\sh}[1]{#1\hskip-7pt \diagup}
\newcommand{\Sh}[1]{#1\hskip-10pt \diagup}

\newcommand{\Eq}[1]{Eq.~(\ref{#1})}

\newcommand{\lqcd}{\Lambda_{\mathrm{QCD}}}
\newcommand{\lqcdsq}{\Lambda^2_{\mathrm{QCD}}}

\definecolor{TGcolor}{rgb}{0.6,0.6,0.0}

\begin{document}

\title{A fresh look at hadronic light-by-light scattering in the muon $g-2$
with the Dyson-Schwinger approach}
\titlerunning{Hadronic light-by-light scattering in the muon $g-2$}

\author{Christian S. Fischer \inst{1,2,3} \and Tobias Goecke \inst{3} \and Richard Williams \inst{3}}
\authorrunning{C. S. Fischer, T. Goecke, R. Williams}

\institute{Institut f\"ur Theoretische Physik, 
 Universit\"at Giessen, 35392 Giessen, Germany 
 \and
  Gesellschaft f\"ur Schwerionenforschung mbH, 
  Planckstr. 1  D-64291 Darmstadt, Germany.
\and 
 Institut f\"ur Kernphysik, 
 Technische Universit\"at Darmstadt, 
 Schlossgartenstra{\ss}e 9, 64289 Darmstadt, Germany}

\abstract{
Using the Dyson-Schwinger and Bethe-Salpeter equations, we calculate the 
hadronic light-by-light scattering contribution to the anomalous magnetic 
moment of the muon $a_\mu$, using a phenomenological model for the gluon 
and quark-gluon interaction. We find $a_\mu=(84 \pm 13)\times 10^{-11}$ 
for meson exchange, and $a_\mu = (107 \pm 2 \pm 46)\times 10^{-11}$ for 
the quark-loop. The former is commensurate with past calculations; the 
latter much larger due to dressing effects. This leads to a revised 
estimate of $a_\mu=116\,591\,865.0(96.6)\times 10^{-11}$, reducing the 
difference between theory and experiment to $\simeq1.9~\sigma$.
}
     
\maketitle

\section{Introduction}

The anomalous magnetic moment $a_\mu = (g_\mu -2)/2$ of the muon is one of
the most precisely determined quantities in particle physics, both
theoretically and experimentally. Their impressive agreement 
on the level of $10^{-8}$ serves as one of the prime examples of the fidelity of 
the Standard Model (SM). Experimental
efforts at Brookhaven and theoretical efforts of the past ten years have
pinned $a_\mu$ down to the $10^{-11}$ level, leading to deviations
between theory \cite{Jegerlehner:2009ry} and experiment \cite{Bennett:2006fi,Roberts:2010cj} 
of about $3$ $\sigma$:
      \begin{align}
		\label{eqn:amuexperiment}
            \mbox{Experiment:} \,\,\,\,
			&116\,592\,089.0(63.0)\times 10^{-11}\;\;, 
		\\
		\label{eqn:amutheoretical}
            \mbox{\phantom{wwu}} \mbox{Theory:} \,\,\,\,
			&116\,591\,790.0(64.6)\times 10^{-11}\;\;.
      \end{align}
This discrepancy is extremely interesting, since it may be a signal for New
Physics beyond the SM. However, to clearly distinguish between New Physics
and possible shortcomings in the SM calculations the uncertainties 
present in both experimental and theoretical values of $a_\mu$ 
need to be further reduced.

\begin{figure}[t!]
		\subfigure[][]{
			\label{fig:hadroniclo}\includegraphics[width=0.25\columnwidth]{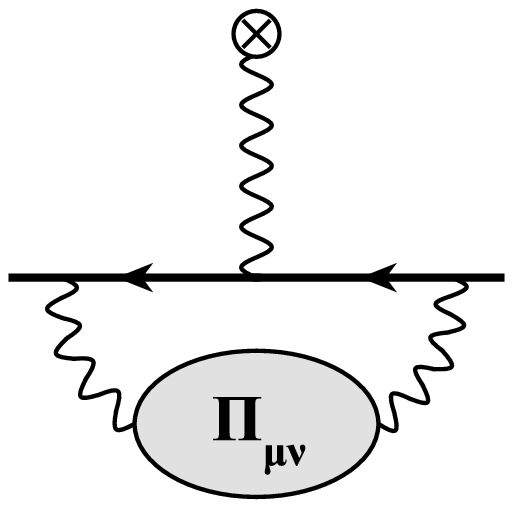}
			}
			\hspace{0.08\columnwidth}
	\subfigure[][]{\label{fig:hadroniclbl}\includegraphics[width=0.25\columnwidth]{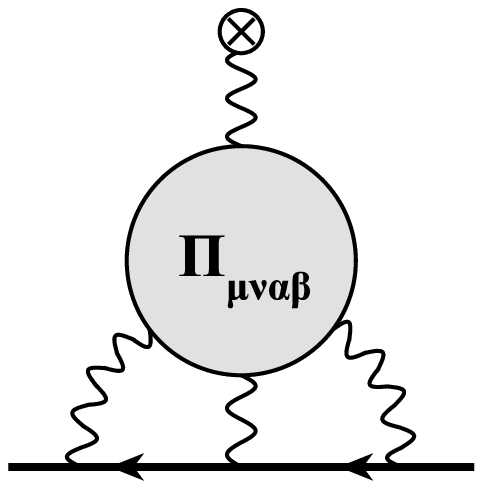}}
      \caption{Two classes of corrections to the photon-muon vertex
	function: (a) Hadronic vacuum polarisation contribution;
	     	   (b) Hadronic light-by-light (LBL) scattering contribution.}
\end{figure}

The greatest uncertainties in the theoretical determination of $a_\mu$
are encountered in the hadronic contributions, i.e.
those terms which involve QCD beyond perturbation theory. The 
most prominent of these is given by the vacuum polarisation tensor 
dressing of the QED vertex, see Fig.~\ref{fig:hadroniclo}. Fortunately it can be
related to experimental data of $e^+ e^-$-annihilation and $\tau$-decay via 
dispersion relations and the optical theorem, resulting in a precise 
determination with systematically improvable errors \cite{Jegerlehner:2009ry}.
Although currently these uncertainties dominate the theoretical error
in \Eq{eqn:amutheoretical} it is foreseeable that 
future experiments reduce this error below that of another, more problematic
source. This is the hadronic light-by-light (LBL) scattering 
diagram, shown in Fig.~\ref{fig:hadroniclbl}. This contribution cannot be directly 
related to experiment and must hence be calculated entirely through theory. 

\begin{figure}[t!]
            \begin{eqnarray*}
            \begin{array}{c}
            \includegraphics[height=1.5cm]{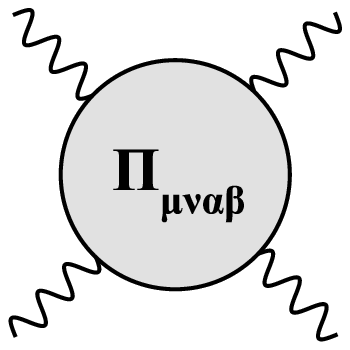}
            \end{array}
            \!\simeq\!
            \begin{array}{c}
            \includegraphics[height=1.5cm]{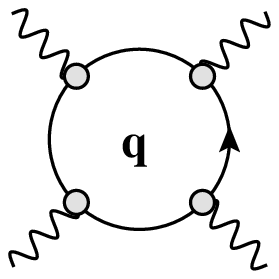}
            \end{array} 
            \!\!+\!\!      
            \begin{array}{c}
            \includegraphics[height=1.5cm]{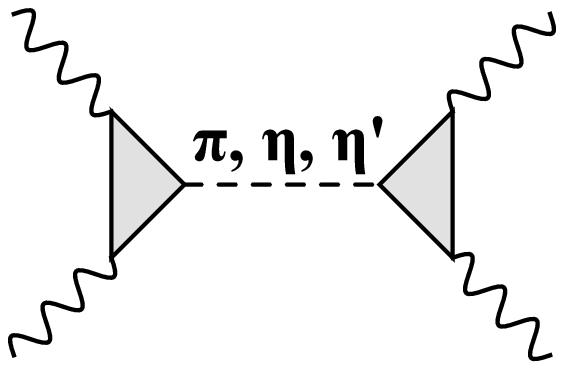}
            \end{array}
            \!\!+\!\!
            \begin{array}{c}
            \includegraphics[height=1.5cm]{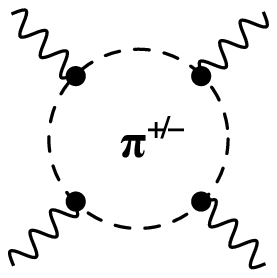}
            \end{array} 
            \end{eqnarray*}
      \caption{The hadronic light-by-light (LBL) contribution to
               $a_\mu$ and its leading expansion in a quark loop (left), 
               pseudoscalar meson exchange (middle) and a
		   meson loop (right). \label{fig:photon4pt}}
\end{figure}

The central object is the photon four-point
function. It receives important contributions from
the small momentum region below $2$~GeV, where perturbative QCD breaks down and 
non-perturbative methods are imperative. In the past the LBL contribution has been 
approximated using ideas from the large-$N_c$ expansion and chiral effective
theories and the associated ordering of diagrams \cite{deRafael:1993za},
see Fig.~\ref{fig:photon4pt}. These diagrams have been 
evaluated within the extended Nambu-Jona-Lasinio model (ENJL) 
\cite{Bijnens:1995cc,Bijnens:1995xf,Bijnens:2007pz}, 
and the hidden local symmetry model \cite{Hayakawa:1995ps,Hayakawa:1996ki,Hayakawa:1997rq}. 
Recent refined calculations
have used ideas of vector meson dominance (VMD) 
\cite{Knecht:2001qf,Melnikov:2003xd,Nyffeler:2009tw,Nyffeler:2010rd}
and a non-local chiral quark model \cite{Dorokhov:2004ze,Dorokhov:2008pw}, 
see~\cite{Prades:2009tw} for a summary. In all these calculations the (pseudoscalar) meson
exchange contributes the most with the meson loop found to be small. An
explanation of the latter is given in \cite{Melnikov:2003xd}.
As a result, we quote the recent value for LBL 
$a_\mu^{\mathrm{LBL}} = 105(26)\times10^{-11}$ proposed in Ref.~\cite{Prades:2009tw}, which
also agrees with~\cite{Nyffeler:2010rd}.

One problem common to all of these approaches has seen intense debate in
recent years~\cite{Jegerlehner:2009ry}: to account for spacelike momenta flowing through
the meson propagator in the exchange diagram one requires a
prescription to take the meson `off-shell'. Such a procedure is not free of ambiguities thus
generating systematic uncertainties which are hard to quantify. 

\begin{figure}[t!]
            \begin{eqnarray*}
            \begin{array}{c}
            \includegraphics[height=1.5cm]{photon4ptfn}
            \end{array}
            \simeq
            \begin{array}{c}
            \includegraphics[height=1.5cm]{photon4ptfn-qrkloop}
            \end{array} 
            +          
            \begin{array}{c}
            \includegraphics[height=1.5cm]{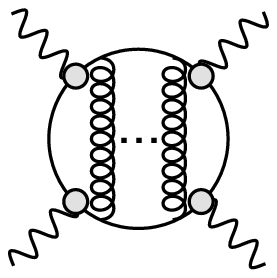}
            \end{array} 
		+
            \begin{array}{c}
            \includegraphics[height=1.5cm]{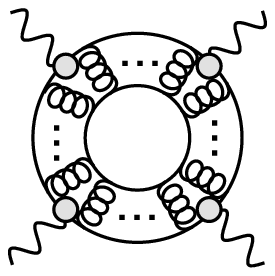}
            \end{array} 
            \end{eqnarray*}
      \caption{The LBL contribution to
               $a_\mu$ and its expansion in a quark loop part (left), 
               a ladder exchange part (middle) and a ladder ring 
		   part (right). All propagators are fully dressed.\label{fig:photon4pt_2}}
\end{figure}

A possibility to avoid such problems is to abandon effective degrees of freedom by
working on the level of (dressed) quarks and gluons. Such a (nonperturbative) 
expansion is given in Fig.~\ref{fig:photon4pt_2}. Here, all propagators are fully dressed, 
with the quark-gluon interaction given in terms of a rainbow-ladder (RL) approximation.
No double-counting is involved.
On the meson mass shells, the ladder sum in the exchange and ring diagrams of 
Fig.~\ref{fig:photon4pt_2} reduce to the corresponding diagrams of Fig.~\ref{fig:photon4pt}.
However, Fig.~\ref{fig:photon4pt_2} accounts systematically and unambiguously for off-shell 
effects which may be more accurate than the effective description of 
Fig.~\ref{fig:photon4pt}. Clearly, the diagrams in 
Fig.~\ref{fig:photon4pt_2} are not the full story and must be supplemented by 
non-RL like diagrams. Nevertheless, they provide a
systematically improvable starting point for a complementary
evaluation of hadronic LBL scattering.

In this letter we briefly report on the first steps in the evaluation of 
Fig.~\ref{fig:photon4pt_2}; a detailed account on our calculation can be 
found in Ref.~\cite{Goecke:2010if}, a very brief summary is given in the proceedings
contribution Ref.~\cite{Goecke:2010sz}.
Our framework for the QCD part of the calculation are the
Dyson-Schwinger (DSE) and Bethe-Salpeter equations (BSE) of Landau gauge QCD 
\cite{Alkofer:2000wg,Maris:2003vk,Fischer:2006ub}.
We present results for the quark loop diagram and for the resonant part of the middle 
diagram of Fig.~\ref{fig:photon4pt_2} in the form of the pseudoscalar meson exchange. 
To our knowledge our calculation is the first to employ fully dressed, momentum dependent
quark propagators and vertices in the quark loop; for the meson exchange diagrams we 
calculate the $\pi \gamma \gamma$ form factors from the underlying theory as opposed 
to the frequently employed strategy of using ans\"atze 
\cite{Knecht:2001qf,Melnikov:2003xd,Nyffeler:2010rd}. As we will see, our results
for the pseudoscalar meson exchange agree nicely with those of previous approaches,
whereas the contribution from the dressed quark loop turns out to be considerably
larger. We discuss these findings at the end of this letter.\\

\section{Calculational scheme: quarks and mesons}

\begin{figure}[b!]
      \includegraphics[width=0.75\columnwidth]{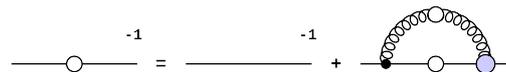}
      \caption{Dyson--Schwinger equation for the quark propagator.}\label{fig:quarkdse}
\end{figure}

Here we summarize our calculational scheme for LBL; the details are
given in Ref.~\cite{Goecke:2010if}. We determine the fully dressed inverse quark propagator, 
	\begin{align}
		S^{-1}(p) = i \sh{p} A(p^2) + B(p^2) \label{quark}
	\end{align}
with vector and scalar dressing functions $A(p^2)$ and $B(p^2)$ from the DSE shown in 
Fig.~\ref{fig:quarkdse}. The Landau gauge gluon propagator $D_{\mu \nu}(k^2)$ is
given by 
	\begin{align}
		D_{\mu\nu}(k) =  \left( \delta_{\mu\nu}-\frac{k_\mu k_\nu}{k^2}  
  			\right)\frac{Z(k^2)}{k^2} 
	\end{align}
with the nonperturbative dressing function $Z(k^2)$. The quark-gluon vertex 
$\Gamma_\mu(p,q)$ with quark momenta $p,q$ in principle consists of contributions 
%
\begin{figure}[b!]
      \includegraphics[width=0.50\columnwidth]{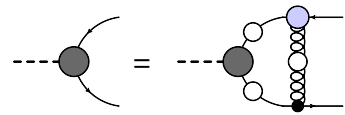}
      \caption{Bethe-Salpeter equation for mesonic bound states.}
      \label{fig:bse}
\end{figure}
from twelve different tensor structures. Here we take the phenomenologically 
important $\gamma_\mu$-part into account leading to \linebreak 
$\Gamma_\mu(p,q) = \Gamma^{\mathrm{YM}}(k^2)\gamma_\mu$ with gluon momentum $k$.
Both functions $Z(k^2)$ and $\Gamma^{\mathrm{YM}}(k^2)$ are combined
into a single model function. For the Maris-Tandy (MT) model~\cite{Maris:1999nt} 
this is
	\begin{eqnarray}
		Z(k^2) \Gamma^{\mathrm{YM}}(k^2) &=&  \frac{4\pi}{g^2} 
      	 \bigg( \frac{\pi}{\omega^6}D q^4 \exp(-q^2/\omega^2)\label{eqn:maristandy}\\
		&&\hspace*{-20mm}+\frac{2\pi \gamma_m}{\log( \tau+(1+q^2/\lqcdsq )^2)}
		 \left[ 1-e^{-q^2/( 4m_t^2 )} \right]\bigg)\;,\nonumber 
	\end{eqnarray}
with $m_t= 0.5\,{\rm GeV}$, $\tau\;=\;\mathrm{e}^2-1$, $\gamma_m=12/(33-2N_f)$, 
$\lqcd\;=\;0.234\,{\rm GeV}$, $\omega=0.4\,{\rm GeV}$ and $D=0.93\,{\rm GeV}^2$. 
The Gaussian factor gives the interaction strength necessary for dynamical 
chiral symmetry breaking, whereas the logarithm represents the one-loop behavior 
of the running coupling at large perturbative momenta. 
The latter is mandatory for the correct short distance behavior
of the quark propagator and all derived quantities.

The rainbow approximation \Eq{eqn:maristandy} leads to the corresponding 
ladder approximation in the kernel of the BSE, Fig.~\ref{fig:bse},
describing mesons as bound states of quarks and antiquarks. 
The Bethe-Salpeter vertex function for a pseudoscalar 
meson is
	\begin{align}\label{eqn:pion}
	  	\Gamma^{q\bar{q}}(p;P)&= \gamma_{5}\left[F_1
               -i\Sh{P} F_2
               -i\sh{p} 
		   F_3
                -\left[\Sh{P},\sh{p}\right]F_4
		    \right]
            \;\; ,
	\end{align}
with $P,p$ the total and the relative momenta of the two
constituents and $F_i:=F_i(p,P)$.

The approximation (Figs.~\ref{fig:quarkdse} and \ref{fig:bse}), with 
the MT model \Eq{eqn:maristandy} is very successful from a phenomenological 
perspective \cite{Maris:1999nt,Maris:1999ta,Maris:2002mz,Maris:2003vk}.
This is especially true for the pseudoscalar meson
sector, wherein the Goldstone-nature of the pion is realized in the chiral limit. While 
tuned to reproduce experimental values for the pion masse and decay constant, it also 
reproduces the pion charge radius and $\pi \gamma \gamma$ transition form factors 
on the percent level. In the vector channel, agreement with experimental masses and decay constants is on the five and ten percent 
level. Thus we view the MT model as a promising starting point for
our calculation of the LBL amplitude.\\ 

\section{The pseudoscalar meson exchange contribution to LBL}

\begin{figure}[b!]
      \includegraphics[width=0.80\columnwidth]{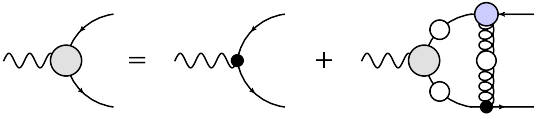}
      \caption{Equation for the quark-photon vertex.}\label{fig:quarkphotondse}
\end{figure}

We begin by calculating the pseudoscalar meson exchange diagram of
Fig.~\ref{fig:photon4pt}, which first requires that we numerically determine
the full quark-photon vertex,
given by its inhomogeneous BSE shown in Fig.~\ref{fig:quarkphotondse}.  
This vertex is non-perturbative and necessary for the calculation of the $\pi\gamma\gamma$ form-factor.
With one Lorentz and two spinor indices, it is
decomposed into twelve Dirac structures; a suitable basis is specified in Ref.~\cite{Ball:1980ay}.
A quark-photon vertex determined thus also contains time-like poles corresponding to vector meson exchange \cite{Maris:1999ta}.
Thus the main ideas of VMD are naturally included here.

\begin{figure}[b!]
      \includegraphics[width=0.27\columnwidth]{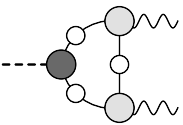}
      \caption{The $\pi^0\gamma\gamma$ form-factor in impulse approximation.}\label{fig:quarktriangle}
\end{figure}

Next we determine the $\pi\rightarrow\gamma\gamma$ form-factor $F^{\pi\gamma\gamma}(k_1,k_2)$,
with the two photon momenta $k_1$ and $k_2$. 
In impulse approximation, consistent with the RL truncation introduced
above, this is given by the diagram in Fig.~\ref{fig:quarktriangle}. 
The $\pi^0$ electromagnetic form-factor has been explored in detail in Ref.
\cite{Maris:2002mz}, wherein it has been confirmed that the correct limit at
vanishing photon momenta, given by the Abelian anomaly, is obtained:
      \begin{align}\label{eqn:scalarformfactor}
           	\Lambda^{\pi\gamma\gamma}_{\mu\nu}(k_1^2,k_2^2) &= 
            i \frac{\alpha_{\mathrm{em}}}{\pi f_\pi} \varepsilon_{\mu\nu\alpha\beta}
            k_1^{\alpha}k_2^{\beta} F^{\pi\gamma\gamma}(k_1^2,k_2^2)\;\;,
      \end{align}
where $\alpha_{\mathrm{em}}$ is the fine structure constant and $f_\pi$ the pion decay 
constant. The prefactors are such that $F^{\pi\gamma\gamma}(0,0)=1$. 
It has been shown analytically and numerically that the form factor has the
correct asymptotic behaviour \cite{Maris:2002mz}: 
\begin{eqnarray}
\lim_{Q^2\rightarrow\infty}F^{\pi \gamma\gamma*}(0,Q^2) &\propto& \frac{1}{Q^2} \,, \nonumber\\ 
\lim_{Q^2\rightarrow\infty}F^{\pi \gamma*\gamma*}(Q^2,Q^2) &\propto& \frac{1}{Q^2} \,.
\end{eqnarray} 
Following the procedure outlined above we obtain the
$\pi\rightarrow\gamma\gamma$ form-factor $F^{\pi\gamma\gamma}(k_1,k_2)$
numerically for all momenta
$k_1$ and $k_2$ necessary to evaluate the meson exchange diagram in LBL,
with a total numerical error of the order of one percent. Similarly, we evaluate the corresponding form 
factors for the $\eta$ and $\eta'$ mesons.

Our solutions for $F^{\pi\gamma\gamma}(k_1,k_2)$ have the same overall structure as the
VMD inspired ans\"atze used in 
Refs.~\cite{Knecht:2001qf,Melnikov:2003xd,Nyffeler:2010rd,Prades:2009tw}.
There are non-negligible differences in the 
mid-momentum region of the order of five percent, whereas the asymptotic
behavior is the same~\cite{Maris:1999ta}. 

Our results for the form factors are used to evaluate the pseudoscalar meson 
exchange contribution to LBL. This requies an off-shell prescription 
for the exchanged mesons. We use the chiral limit axial-vector Ward-Takahashi 
identity
	\begin{align}\label{eqn:chiralaxwti}
		2P_\mu \Gamma^{5}_\mu(k,P) = i S^{-1}(k_+)\gamma_5 + i\gamma_5 S^{-1}(k_-)\;\;.
	\end{align}
that establishes the following form for the dominant $\pi^0$-amplitude:
	\begin{align}\label{eqn:offshellprescriptionScalar}
 		F_1(k,P)=\lambda_3\left(B(k_+)+B(k_-)\right)/\left(2f_\pi\right)\;,
	\end{align}
with $k_{\pm} = k\pm P/2$ and $\lambda_3$ a Gell-Mann matrix. On mass-shell in
the chiral limit
it reduces to the exact result $F_1(k,0) :=\lambda_3 B(k^2)/f_\pi$.
The pseudoscalar amplitude of Eq.~(\ref{eqn:pion}) is generalized via
Eq.~(\ref{eqn:offshellprescriptionScalar}) for all Dirac structures.

The systematic error of our calculation can be attributed entirely to the validity
of the RL approximation within the MT model, \Eq{eqn:maristandy}, and the
off-shell prescription \Eq{eqn:offshellprescriptionScalar}. No other approximations 
have been used. While in the Goldstone-Boson sector the MT model works well, there 
is certainly a larger error in the flavor singlet sector. We therefore
guesstimate a total systematic error:
ten percent for the pion contribution, and twenty percent for the $\eta$ and $\eta'$
contributions. With a numerical error of two percent we 
obtain:
$a_\mu^{\textrm{LBL};\pi^0}=(57.5 \pm 6.9)\times 10^{-11}$,
$a_\mu^{\textrm{LBL};\eta} =(15.8 \pm 3.5)\times 10^{-11}$ and
$a_\mu^{\textrm{LBL};\eta'}=(11.0 \pm 2.4)\times 10^{-11}$ leading to
	\begin{align}
		a_\mu^{\textrm{LBL;PS}}=(84.3 \pm 12.8)\times 10^{-11} \label{res:PS}
	\end{align}
This number is
compatible with previous results 
\cite{Nyffeler:2009tw,Nyffeler:2010rd,Prades:2009tw,Dorokhov:2008pw}.\\

\section{Quark loop contribution to LBL}

We now focus on the quark loop
contribution to LBL. Here we follow the strategy employed in 
Ref.~\cite{Aldins:1969jz,Aldins:1970id}, and differentiate the four-point function 
wrt. the external photon momentum~\cite{Bijnens:1995cc}. We verified that 
we reproduce the well known perturbative result for the electron loop
with an accuracy of better than one per mille. For the quark loop we use 
the fully dressed quark propagators for the up, down, strange and charm 
quarks, extracted from their DSE, Fig.~\ref{fig:quarkdse}. For the 
quark-photon vertex one would like to use the full numerical
solution of Fig.~\ref{fig:quarkphotondse}, which satisfies the Ward-Identity
of the vertex and contains transverse parts including vector meson contributions.
However, this is a formidable numerical task which is beyond the scope of the
present work. Instead, we concentrate on that part of the vertex which is 
constrained by gauge invariance and compare the result for bare vertices 
with one where we use the first term of the Ball-Chiu representation 
(1BC)~\cite{Ball:1980ay}
	\begin{align}
		\Gamma_\mu(p,q) = \gamma_\mu \,\left(A(p^2)+A(q^2)\right)/2 \,,\; \label{vertexdressing}
	\end{align}
where $p,q$ are the quark and antiquark momenta. This expression is known to be a 
reasonable approximation of the full Ball-Chiu vertex, which is constructed to 
satisfy the Ward-Takahasi identity (WTI) of the complete quark-photon vertex. Thus
one may hope that Eq.~(\ref{vertexdressing}) gives a good first estimate for those
parts of the vertex relevant for gauge invariance. Nevertheless, one has to keep 
in mind that potentially important transverse contributions due to vector meson
poles \cite{Bijnens:1995cc,Bijnens:1995xf} are neglected. Possible quantitative implications of 
this omission are discussed in \cite{Goecke:2010if}.

Here, comparing between the bare result and that with \Eq{vertexdressing} may serve as a 
first guide for the systematic error due to the relevance of vertex effects. 
We find
	\begin{equation}
		\begin{array}{lcc}
		a_\mu^{\textrm{LBL;quarkloop (bare vertex)}} &=&
		(\phantom{0}61 \pm 2) \times 10^{-11}\\
		a_\mu^{\textrm{LBL;quarkloop (1BC)}}         &=& 
		(107 \pm 2)		    \times 10^{-11}\\
		\end{array}\label{res:QL} \\
	\end{equation}
for the quark loop contribution. Clearly this is a sizable contribution. Whereas the
bare vertex result roughly agrees with the number $60 \times 10^{-11}$ given in 
\cite{Melnikov:2003xd}, the dressing effects of the vertex lead to a drastic 
increase. Unfortunately this makes it very hard if not impossible
to guess the effect of the total vertex dressing without an explicit calculation. 
Certainly given these findings, all previously given estimates for the 
systematic error in the quark loop contributions seem to be an order of magnitude too small.\\

\section{Conclusions}

In this letter we have presented a new approach towards the anomalous magnetic moment of
the muon. We have used a combination of Dyson-Schwinger and Bethe-Salpeter equations
to evaluate the pseudoscalar meson exchange contribution and the quark loop contribution
to LBL. Our only input is the Maris-Tandy model, a phenomenologically successful ansatz 
for the combined strength of the gluon propagator and the quark-gluon vertex. We believe 
this is a systematic improvement as compared to previous approaches. When combining our two
results, \Eq{res:PS} and \Eq{res:QL}, we also have to add the 
effects from the right hand diagrams of Fig.~\ref{fig:photon4pt} or Fig.~\ref{fig:photon4pt_2}. 
These are contributions involving an additional quark-loop which are typically
negative and of the order of ten to twenty percent of the leading-$N_c$ contributions 
\cite{Fischer:2007ze,Fischer:2008wy,Fischer:2009jm}. Since on the other hand we also expect positive 
contributions of a similar size from non-pseudoscalar exchange diagrams \cite{Jegerlehner:2009ry}
we choose to subsume all these contributions to another 
$a_\mu^{\textrm{LBL;other}}=(0 \pm 20)\times 10^{-11}$, where the error is clearly
subjective. This amounts to a total error so far of $35 \times 10^{-11}$. Because of the 
importance of dressing effects, we use the quark-loop in the 1BC approximation for the 
central value and gauge an additional systematic error of $46 \times 10^{-11}$
by comparing with the bare vertex result. This gives us the following total hadronic LBL
contribution 
	\begin{equation}
		\begin{array}{lcc}
		a_\mu^{\textrm{LBL; (1BC)}}         &=& (191 \pm 35 \pm
		46)\times 10^{-11}\;, \\
		\end{array}
	\end{equation}
in our approach.
Taken at face value these numbers together with the other contributions
quoted in \cite{Jegerlehner:2009ry} clearly reduce the discrepancy between theory and 
experiment. Combining our light-by-light scattering results with the
other SM contributions gives:
	\begin{align}\label{eqn:newamu}
		a_\mu^{\textrm{theor.}}=116\,591\,865.0(96.6)\times 10^{-11}\;\;.
	\end{align}
With the positive shift and larger error, the deviation between theory
and experiment is reduced to $\simeq1.9~\sigma$.

Whether the neglected parts of the fermion-photon vertex have the potential 
to drive this theoretical result further towards the experiment or back to
previous results \cite{Bijnens:2007pz,Jegerlehner:2009ry} remains at present a 
subject of mere speculation. Certainly, our results do not provide final answers in any 
sense but may serve as a starting point towards a more fundamental determination of 
$a_\mu$ along the lines discussed around Fig.~\ref{fig:photon4pt_2}.

What we have shown, however, is that dressing effects in the quark-photon vertex
due to gauge invariance constraints are important and cannot be ignored in 
determining reliable estimates of hadronic LBL.\\

{\bf Acknowledgments}\\
We are grateful to Andreas Krassnigg
for fruitful interactions in the early stages of this work. This work was supported 
by the Helmholtz-University Young Investigator Grant No.~VH-NG-332 and by the 
Helmholtz International Center for FAIR within the LOEWE program of the State of Hesse.

\end{document}